\documentclass[%
 reprint,
 amsmath,amssymb,
 aps,
 superscriptaddress
]{revtex4-1}

\usepackage{graphicx}
\usepackage{dcolumn}
\usepackage{bm}
\usepackage{hyperref}
\hypersetup{
    colorlinks,
    citecolor=black,
    filecolor=black,
    linkcolor=black,
    urlcolor=black
}
\usepackage{bbm}
\usepackage{float}
\usepackage{braket}
\usepackage{enumerate}

\begin{document}

\newcommand{\bb}[1]{{\color{blue} BJB: #1}}

\newcommand{\mk}[1]{{\color{red} MSK: #1}}

\newcommand{\comment}[1]{}

\newcommand{\ft}[1]{{\color{red} FT: #1}}

\newcommand{\sdb}[1]{{\color{green} SDB: #1}}

\newcommand{\addition}[1]{{\color{blue} #1}}

\title{Low-overhead quantum computing with the color code}

\author{Felix~Thomsen}
\affiliation{Centre for Engineered Quantum Systems, School of Physics,
University of Sydney, New South Wales 2006, Australia}
\author{Markus~S.~Kesselring}
\affiliation{Dahlem Center for Complex Quantum Systems, Freie Universität Berlin, 14195 Berlin, Germany}
\author{Stephen~D.~Bartlett}
\affiliation{Centre for Engineered Quantum Systems, School of Physics,
University of Sydney, New South Wales 2006, Australia}
\author{Benjamin~J.~Brown}
\affiliation{Centre for Engineered Quantum Systems, School of Physics,
University of Sydney, New South Wales 2006, Australia}

\date{\today}

\begin{abstract}
Fault-tolerant quantum computation demands significant resources:  large  numbers  of  physical qubits must be checked for errors repeatedly to protect quantum data as logic gates are implemented in the
presence of noise. We demonstrate that an approach based on the color code can lead to considerable reductions in the resource overheads compared with conventional methods, while remaining compatible with a two-dimensional layout. We propose a lattice surgery scheme that exploits the rich structure of the color-code phase to perform arbitrary pairs of commuting logical Pauli measurements in parallel while keeping the space cost low. 
Compared to lattice surgery schemes based on the surface code with the same code distance, our approach yields about a $3\times$ improvement in the space-time overhead, obtained from a combination of a $1.5\times$ improvement in spatial overhead together with a $2\times$ speedup due to the parallelisation of commuting logical measurements. Even when taking into account the color code's lower error threshold using current decoders, the overhead is reduced by 10\% at a physical error rate of $10^{-3}$ and by 50\% at $10^{-4}$. 

\end{abstract}

\maketitle

\section{Introduction}

A quantum computer capable of executing large calculations will likely require a fault-tolerant design, employing quantum error correction to encode logical qubits in many physical qubits and to perform logical gates on the encoded information~\cite{1998preskill,2003kitaev,Dennis02,2015terhal, 2017roadstowards}. 
Furthermore, the design should respect the layout of the physical qubits of the available hardware. For instance, many solid-state quantum devices have two-dimensional nearest-neighbour connectivity~\cite{google2022, Krinner2022, zhao2022, Sundaresan2023}.
Ideally, a practical fault-tolerant architecture should be resource-efficient, executing quantum algorithms of interest with a low overhead in terms of the numbers of both additional physical qubits and the time needed to perform logic gates.

The color code~\cite{2006bombin} is particularly well suited for fault-tolerant quantum computing using a two-dimensional qubit array. It has low-weight stabilizer checks that can be read out using a nearest-neighbour qubit layout~\cite{Chamberland_2020, beverland2021cost} and it has a fault-tolerant implementation of logical Clifford gates by transversal rotations on its physical qubits. In addition to these inherently fault-tolerant transversal gates, the color code also admits a multitude of fault-tolerant measurement-based code deformations~\cite{Raussendorf07fault, Bombin09CodeDefo,2011fowlercolor, Horsman_2012, 2017benpoking}. The wide variety of available fault-tolerant logic gates can be attributed to the rich structure of its underlying anyon model when viewed as a topological phase~\cite{Yoshida15,Bridgeman17, Kesselring_2018}.

In spite of its practicality, and its favourable properties for performing logical operations, little work has been conducted to determine the resource cost of performing large quantum algorithms with the color code realised on a two-dimensional qubit array.

In this manuscript we propose a scheme for fault-tolerant quantum computing based on the color code~\cite{2006bombin} where we perform logical operations using an ancillary lattice of qubits as a resource to make Pauli measurements between logical data qubits and sources of magic states~\cite{Bravyi05,Litinski_2019, chamberland2020building, chamberland2021universal, cohen2021lowoverhead} using lattice surgery~\cite{Horsman_2012, landahl2014quantum, 2017benpoking, 2018litinskitwist, cohen2021lowoverhead}. This scheme, compatible with a two-dimensional architecture, offers a substantial reduction in the physical resource cost of fault-tolerant quantum computing against the surface code---the archetype for fault-tolerant quantum computing~\cite{2012surface, Litinski_2019,2019litinskinotascostly, fowler2019low,2019gidneymagic}. 
Compared with the most resource efficient mode of surface-code quantum computation~\cite{Litinski_2019}, our scheme reduces the space-time overhead as a function of the code distance by approximately a factor of $3$. Even accounting for the color code's lower error threshold using existing decoders, the space-time overhead is reduced by 10\% at a physical error rate of $10^{-3}$ and 50\% at $10^{-4}$. These results elevate consideration of the color-code architecture as a promising pathway for large-scale fault-tolerant quantum computers and motivate further work towards better color-code decoders~\cite{kubica2019efficient, sahay2021decoder, gidney2023new,zhang2023facilitating} and syndrome extraction circuits~\cite{2011fowlercolor, landahl2011faulttolerant,stephens2014efficient, Chamberland_2020, beverland2021cost, bombin2023unifying, McEwen_2023}.

\section{The color code}

\begin{figure}
\includegraphics[width=0.85\columnwidth]{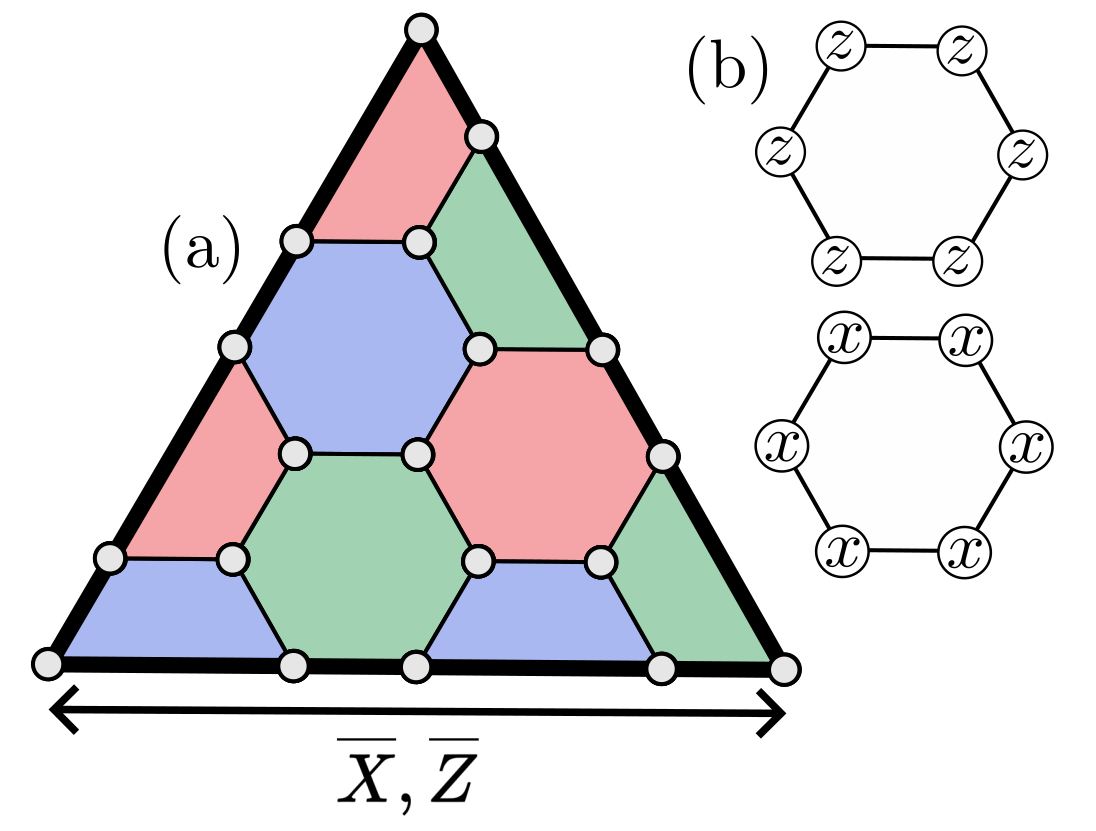}
\caption{(a) Triangular color code with one boundary of each color, encoding one logical qubit. Physical qubits live on the vertices. Each boundary supports both logical $\overline{X}$ and $\overline{Z}$ operators. (b) Each face hosts both $X$- and $Z$-type stabilizers. }
 \label{Fig:triangle}
\end{figure}

The color code~\cite{2006bombin} is defined on a three-colorable lattice where faces are given a color, red, green or blue, such that no two adjacent faces have the same color. A qubit is placed on each vertex, indexed with labels $v$, with a Pauli-$X$ type stabilizer and a Pauli-$Z$ type stabilizer on each face, with indices $f$. The encoded subspace, spanned by code states $|\psi \rangle$, is such that $ S | \psi \rangle = | \psi \rangle $ for commuting stabilizer operators $S$~\cite{Gottesman97}. For the color code we have two types of stabilizers, Pauli-$X$- and Pauli-$Z$-type stabilizers 
\begin{equation}
S^X_f = \prod_{v\in \partial f} X_v, \quad S^Z_f = \prod_{v\in \partial f} Z_v,
\end{equation}
respectively, where $X_v$ and $Z_v$ are standard Pauli matrices acting on vertices $v$ and $\partial f$ denotes the vertices lying on face $f$. We assign each lattice boundary one of three colors, red, green or blue, such that the vertices of lying at boundary of a given color support no faces of that color. We note that some boundary vertices only support a face of a single color. We say that these vertices lie at a corner of the boundary. With our definition, these corner vertices support two boundaries of different color. We can think of these corners as locations where the color of the lattice boundary changes.

In our scheme, we encode logical data qubits using copies of the triangular color code with one boundary of each color~\cite{2006bombin}; see Fig.~\ref{Fig:triangle} where we show a single copy of the code. We index the logical data qubits with indices $1 \le j \le N$. We write the stabilizer group of the data qubits as $\mathcal{S}_\textrm{data}$. This encoding has a transversal implementation of all single-qubit Clifford gates~\cite{2006bombin} and, moreover, all of its logical Pauli operators can be supported entirely on any one of its three boundaries. Specifically, we can write down logical operators
\begin{equation}
\overline{X}_j = \prod_{v \in j, \, \textrm{red bdry.}} X_v, \quad \overline{Z}_j = \prod_{v \in j, \, \textrm{red bdry.}} Z_v,
\end{equation}
where we take the product over all of the vertices of data qubit $j$ supported on the red boundary.
The latter property is favorable for quantum computing by lattice surgery~\cite{Litinski_2019}.

\section{Color code lattice surgery scheme}

\subsection{Overview}

We perform logic gates by making fault-tolerant logical Pauli measurements on the triangular color codes, implemented using an ancillary resource comprised of a lattice of qubits with its boundary aligned with one of the three-colored boundaries of each of the data qubits.

Algorithms proceed via a sequence of these non-destructive logical Pauli measurements between data qubits and encoded magic states~\cite{Litinski_2019}. Here we show that we can measure an arbitrary pair of commuting logical Pauli operators, $\overline{L}_A$ and $\overline{L}_B$, in parallel. In contrast, schemes using the surface code can only make a single arbitrary logical measurement at a time. At a high-level, this improvement can be explained by interpreting the color code as two copies of the surface code~\cite{Bombin_2012, Kubica_2015, Criger_2016, Kesselring_2018}. Thus, the color code has a larger capacity to perform logical operations in parallel. This gives our scheme an advantage in terms of the temporal resources.

The logical Pauli measurements are performed using merging and splitting operations~\cite{Horsman_2012, landahl2014quantum} with an appropriately prepared ancilla system. 

We begin in the stabilizer group $\mathcal{S} = \mathcal{S}_{\textrm{data}} \otimes \mathcal{S}_{\textrm{anc}}$, where $\mathcal{S}_{\textrm{data}}$ is the usual color-code stabilizer group such that we have a single triangular color code for each data qubit, and $\mathcal{S}_{\textrm{anc}}$ is the stabilizer group for the ancilla (details below).

During the merging operation, we measure local generators of a new stabilizer group $\mathcal{S}_M$ which connect the ancilla system to the data qubits. Measuring $\mathcal{S}_M$ lets us infer the values of two operators $Q_{A}, Q_{B} \in \mathcal{S}_M$ that we will use to determine the outcomes of our desired logical measurements $\overline{L}_{A/B}$.

The merging operation is followed by the splitting operation, wherein we read out the ancilla system and recover the codespace of the data qubits by measuring the stabilizer group $\mathcal{S}$ again. In doing so, we also obtain the values of operators $S_A,\,S_B \in  \mathcal{S}$ that are supported on the ancilla system such that $\overline{L}_{A/B} = S_{A/B}Q_{A/B}$. The operators $Q_{A/B}$, which we measured during the merging operation, commute with all elements of $\mathcal{S}$ and so we achieve the desired measurement outcomes. It is worth pointing out that $S_A,\,S_B \not\in \mathcal{S}_M$ and so their measurement during the splitting operation is central to our ability to infer $\overline{L}_{A/B}$, which is a point of difference between our scheme and those based on surface code lattice surgery.

In what follows, we will first discuss how to construct a color code ancilla system with a suitable configuration of boundaries to facilitate an arbitrary pair of commuting measurements $\overline{L}_{A/B}$ in Sec. \ref{construction}. We then describe the merging operation and discuss the aforementioned $Q_{A/B}$ and $S_{A/B}$ operators arising from our construction, which are required to infer $\overline{L}_{A/B}$, in Sec. \ref{merge}. Next, in Sec.~\ref{split}, we describe the splitting operation, confirming that we measure no logical information other than the values of $\overline{L}_{A/B}$. We then provide a comprehensive example that illustrates every step of our scheme in Sec. \ref{example} and discuss the fault-tolerance of these operations in Sec.~\ref{fault-tolerance}.

\subsection{Ancilla system construction} \label{construction}

\begin{figure*}
\includegraphics[width=1.0\textwidth]{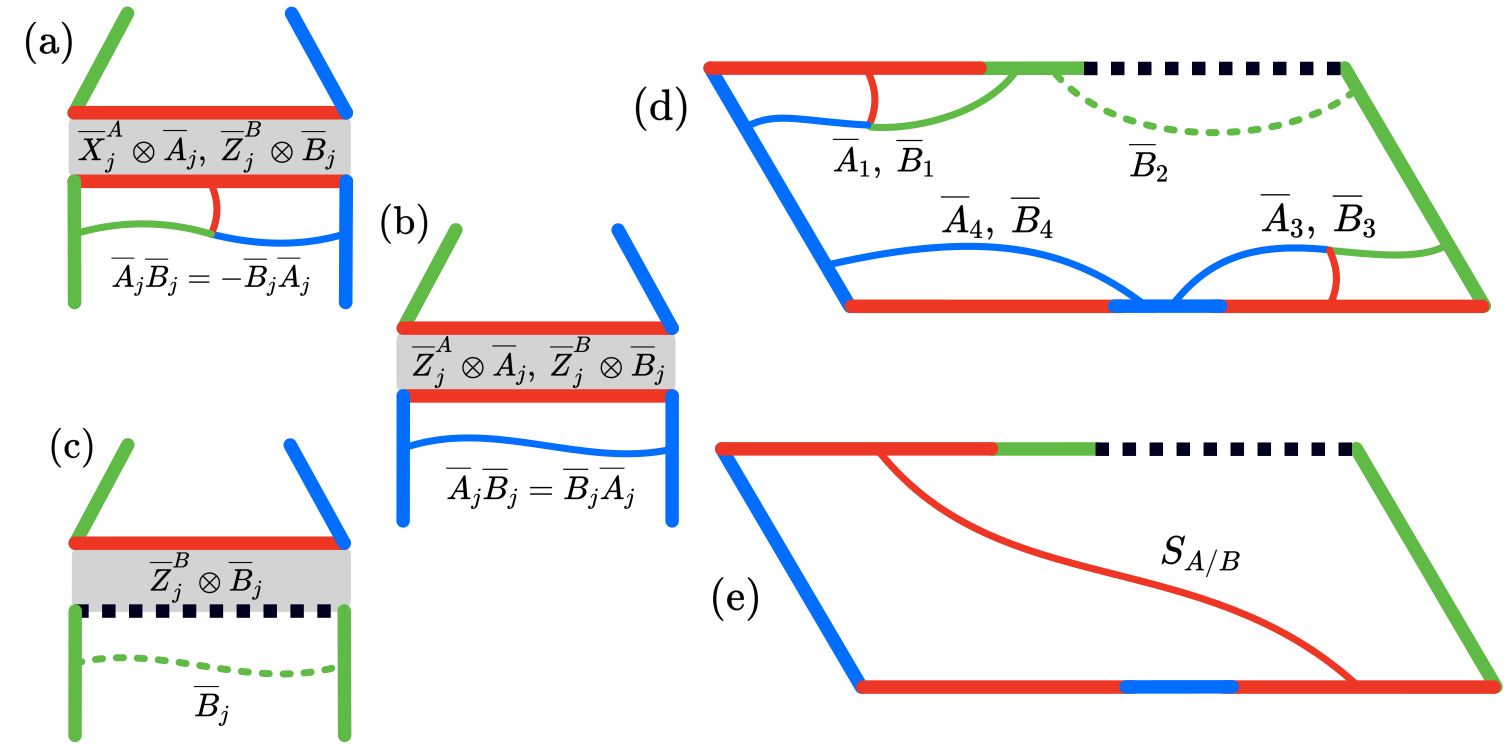}
\caption{Boundaries of the ancilla system. (a)-(c) For a given pair of measurements $\overline{L}_{A/B}$, we assign an ancilla system boundary with logical operators $\overline{A}_j$ and $\overline{B}_j$, which are the products of Pauli-$X$ and Pauli-$Z$ operators along the boundary, for each connection with a data qubit $j$ that is non-trivially acted on by $\overline{L}_{A/B}$. We choose these boundaries to be red if qubit $j$ is acted on non-trivially by both $\overline{L}_{A/B}$ and of Pauli-$Z$ ($X$) type if it is acted on non-trivially only by $\overline{L}_{A}$ ($\overline{L}_{B}$). The commutation relations of $\overline{P}^{A}_j$ and $\overline{P}^{B}_j$ then dictate the required commutation relations of $\overline{A}_j$ and $\overline{B}_j$, such that the parity measurements $\overline{P}^{A}_j \otimes \overline{A}_j$ and $\overline{P}^{B}_j \otimes \overline{B}_j$ obtained during the merging operation commute. We show the required macroscopic structure of $\overline{A}_j$ and $\overline{B}_j$ and the resulting configuration of the adjacent boundaries for examples of the three non-trivial cases 1, 2, 3 discussed in the main text in (a), (b) and (c) respectively. (d) Making a choice for the color of one of the non-connecting boundaries then completely determine the types of all the boundaries of the ancilla system. Here we show an example boundary configuration for an ancilla system constructed for the two measurements $ \overline{L}_A = \overline{X}_1 \overline{X}_3 \overline{Z}_4$ and $ \overline{L}_B = \overline{Z}_1 \overline{Z}_2 \overline{Z}_3  \overline{Z}_4$.  (e) The products of the ancilla logical operators,  $\prod_{j=1}^N \overline{A}_j$ and $\prod_{j=1}^N \overline{B}_j$, are equivalent, up to multiplication by stabilizers, to the operators $S_{A,/B}$, which are a subset of the red string operators stretching between pairs of red boundaries connecting the ancilla system with qubits involved in $\overline{L}_{A/B}$ as in case 1; see the main text for details. These operators are central to our ability to measure the values of $\overline{L}_{A/B}$.}
 \label{Fig:macroscopics}
\end{figure*}

We now describe our protocol for constructing a color code ancilla system with a suitable boundary configuration to facilitate an arbitrary pair of commuting measurements $\overline{L}_{A/B}$. While strictly speaking we never need to prepare the ancilla system in isolation, to describe the logical measurement, we find it helpful to describe the ancilla system as a disjoint code in and of itself. In practice, we can prepare the ancilla system and connect it to the data qubits in parallel when we deform onto the stabilizer group $\mathcal{S}_M$. In what follows, we will, however, view the ancilla system as a code, that encodes its own ancillary logical qubits.  This will help us understand the logical action of the merge and split operations on the logical data qubits.

We choose different boundary conditions for the ancilla system depending on the logical operators we choose to measure. 
Let us write the two commuting Pauli operators of the logical data qubits we want to measure as 
\begin{equation}
\overline{L}_{A/B}= \prod_{j=1}^N \overline{P}^{A/B}_j, 
\end{equation}
where $\overline{P}^{A/B}_j \in \{ \openone, \,\overline{X}_j, \overline{Y}_j, \overline{Z}_j \}$ is the action of the logical operator $\overline{L}_{A/B}$ on data qubit $1 \le j \le N$. For each logical qubit $j$ there are then three non-trivial cases to consider:
\begin{enumerate}
\item $\overline{P}^A_j $ and $ \overline{P}^B_j $ anti commute, \label{item:antiCommutingPair}

\item $\overline{P}^A_j =  \overline{P}^B_j \not= \openone$, \label{item:commutingPair}

\item $\overline{P}^A_j \not=  \overline{P}^B_j $ and one of either $ \overline{P}^{A/B}_j = \openone$. \label{item:measureOne}

\end{enumerate}

The boundary of the ancilla system that lies in the vicinity of a data qubit that is non-trivially acted on by $\overline{L}_{A/B}$, say data qubit $j$, is connected to said data qubit via measurements during the merging step. As such, let us refer to the boundary of the ancilla system that locally neighbours a data qubit as a connecting boundary. We can index these connecting boundaries $j$ according to the index of the data qubit they neighbour.

Each data qubit $j$ has corresponding logical operators supported on the associated connecting boundary of the ancilla system. Let us denote the ancilla logical operators made up of Pauli-$X$ and Pauli-$Z$ operators along the connecting boundaries of the ancilla system assigned to data qubit $j$ as $\overline{A}_j$ and $\overline{B}_j$ respectively.

The boundary type for each connecting boundary $j$ is determined by the form of $\overline{P}^A_j $ and $ \overline{P}^B_j $. We assign the boundary of the ancilla system that neighbours data qubit $j$ to be red if the data qubit is non-trivially involved in both measurements, i.e. $\overline{P}^A_j \not= \openone$ and $ \overline{P}^B_j \not= \openone$, i.e., cases and {\ref{item:antiCommutingPair}} and \ref{item:commutingPair},  and assign it to be a Pauli-$Z$ ($X$) boundary \cite{Kesselring_2018} if only $\overline{P}^A_j \not= \openone$ ($\overline{P}^B_j \not= \openone$); case \ref{item:measureOne}.

We must also specify the boundaries in between the connecting boundaries. These boundaries are coloured either blue or green where, again, their boundary type is determined by the choice of operators $\overline{L}_A$ and $\overline{L}_B$. Let us concentrate on the boundary type of the two boundaries that are adjacent to the connecting boundary $j$. Specifically, these are the boundary that lies in between connecting boundary $j-1$ and $j$, and the boundary that lies in between connecting boundary $j$ and $j+1$.

During the merging operation of our protocol we will measure local stabilizer generators in $\mathcal{S}_M$ connecting boundary $j$ to its respective data qubit to fault-tolerantly infer the parities $\overline{P}^{A}_j \otimes \overline{A}_j$ and $\overline{P}^{B}_j \otimes \overline{B}_j$. The colours of the adjacent boundaries must then be chosen such that $\overline{A}_j$ and $\overline{B}_j$ satisfy the required commutation relations to make these two parity measurements commute, i.e. $[\overline{P}^{A}_j \otimes \overline{A}_j, \overline{P}^{B}_j \otimes \overline{B}_j]=0$.

For cases \ref{item:commutingPair} and \ref{item:measureOne} itemised above, where $\overline{P}_j^A$ and $\overline{P}_j^B$ commute and we therefore require $\overline{A}_j$ and $\overline{B}_j$ to commute also, we find that both of these boundaries have the same colour; either both boundaries are green or both boundaries are blue.

In contrast, in case \ref{item:antiCommutingPair} where $\overline{P}_j^A$ and $\overline{P}_j^B$ anti commute and we require $\overline{A}_j$ and $\overline{B}_j$ to anti commute, the two boundaries adjacent to connecting boundary $j$ must have different colours; one must be blue and one must be green.

At a microscopic level, these rules can be understood by considering the number of qubits along connecting boundary $j$, and therefore in the support of $\overline{A}_j$ and $\overline{B}_j$. Since $\overline{A}_j$ and $\overline{B}_j$ are made up of Pauli-$X$ and Pauli-$Z$ terms respectively, we require connecting boundary $j$ to have to have an even number of qubits for $\overline{A}_j$ and $\overline{B}_j$ to commute and an odd number for them to anti-commute.

It is worth noting here that assigning a Pauli-$X$ ($Z$) instead of a red boundary in case \ref{item:measureOne} to data qubit $j$ has the effect of turning $\overline{A}_j$ ($\overline{B}_j$) into a stabilizer of the ancilla system, i.e. in this case $\overline{A}_j (\overline{B}_j)=\openone$ by construction.

In the color code, a boundary with an odd (even) number of qubits will have adjacent boundaries of different (same) type. This is perhaps most easily understood by considering the macroscopic structure of the string-like logical operators associated with a given boundary configuration and observing the number of crossings. We illustrate the structure of the macroscopic logical operators associated with cases \ref{item:antiCommutingPair}-\ref{item:measureOne} in Fig.~\ref{Fig:macroscopics} (a)-(c).

Given that the boundary type of a segment of the boundary of the ancilla system depends on the connection of the ancilla system to two different data qubits, one may worry whether or not it is possible to globally colour the boundary of the ancilla system in a consistent way. Interestingly, one can check that a global colouring is possible if and only if $\overline{L}_A$ and $\overline{L}_B$ commute. By choosing one of the non-connecting boundaries to be either green or blue, we can then completely determine the types of all the remaining boundaries of the ancilla system. Following this procedure allows for the construction of a suitable ancilla system for any given pair of commuting measurements $\overline{L}_{A/B}$.

Let us also introduce another type of logical operator on the ancilla system; the red-string logical operators $\overline{R}^{X}_{j,k}$ and $\overline{R}^{Z}_{j,k}$. These are string operators made up of Pauli-$X$ and Pauli-$Z$ terms, respectively, supported on red edges of the color-code lattice that run through the bulk of the ancilla system, and terminate at connecting boundaries $j$ and $k$. We note that all of the red string operators are mutually commuting, i.e., $\left[ \overline{R}^{X}_{j,k}, \overline{R}^{X}_{l,m} \right] = \left[ \overline{R}^{Z}_{j,k}, \overline{R}^{Z}_{l,m} \right] = \left[ \overline{R}^{X}_{j,k}, \overline{R}^{Z}_{l,m} \right] = 0$ for any choice of $j, k, l, m$. We also have that $\overline{A}_j$ and $\overline{A}_k$ ($\overline{B}_j$ and $\overline{B}_k$) will anti commute with $\overline{R}^{Z}_{j,k}$ ($\overline{R}^{X}_{j,k}$).

We initialise the system in an eigenstate of all of the red string logical operators by initialising every pair of qubits supported on a red edge in a Bell pair, before measuring the stabilizers of the ancilla system.
Initialising the ancilla system in a fixed state that is an eigenstate of all $\overline{R}^{X}_{j,k}$ and $\overline{R}^{Z}_{j,k}$  therefore ensures that the outcomes of the parity measurements $\overline{P}^{A}_j \otimes \overline{A}_j$ and $\overline{P}^{B}_j \otimes \overline{B}_j$ that are obtained during the merging operation will be individually random and we will not obtain any information about the state of individual data qubits.

\subsection{Merging operation} \label{merge}

During the merging operation we measure local generators of a new stabilizer group, $\mathcal{S}_M$, which includes stabilizers connecting data qubits to their assigned ancilla system boundaries.  

This new stabilizer group allows us to infer the values of global operators 
\begin{equation}
    Q_{A} = V_{A} \prod_{j=1}^N \overline{P}^{A}_j \otimes \overline{A}_j = \overline{L}_{A} S_{A}, \label{Eqn:QoppA}
\end{equation}
\begin{equation}
    Q_{B} = V_{B} \prod_{j=1}^N \overline{P}^{B}_j \otimes \overline{B}_j = \overline{L}_{B} S_{B}, \label{Eqn:QoppB}
\end{equation}
where $\prod_{j=1}^N \overline{P}^{A}_j \otimes \overline{A}_j$ and $\prod_{j=1}^N \overline{P}^{B}_j \otimes \overline{B}_j$ are the products of the local parity measurements between data and ancilla logical operators and $V_{A/B}$ is a stabilizer in $S_M$. 

In the general case, we can choose to take 
\begin{equation}
V_{A} = \prod_{f:\textrm{red,blue}} S_f^{X},\quad V_{B} = \prod_{f:\textrm{red,blue}} S_f^{Z},
\end{equation} 
where the product is taken over red and blue stabilizer generators supported entirely on the ancilla system. With this choice one can check that 
\begin{equation}
V_{A} = S_{A} \prod_{j=1}^N \overline{A}_j, \quad V_{B} = S_{B} \prod_{j=1}^N \overline{B}_j,
\end{equation}
where $S_{A}$ and $S_{B}$ are the products of red Pauli-$X$ and Pauli-$Z$ strings, respectively, supported on all blue boundaries and along a subset of the connecting boundaries involved in $\overline{L}_{A/B}$ as in cases \ref{item:commutingPair} and \ref{item:measureOne} on the ancilla system. We show macroscopic and microscopic examples of these operators in Figs.~\ref{Fig:macroscopics} and \ref{Fig:Architecture} respectively, but note this holds in the general case. Indeed, as the example shown in the figures includes all three cases we have defined above, we can regard it as as exhaustive.
With this choice of $V_{A/B}$ we obtain the right-hand side of the results shown in Eqns.~(\ref{Eqn:QoppA}) and~(\ref{Eqn:QoppB}).

Note that while $S_{A/B} \in \mathcal{S}_{\textrm{anc}}$, the measurements we perform during the merging operation anti-commute with $S_{A/B}$ and so $S_{A/B} \not\in \mathcal{S}_M$. Measuring these operators during the splitting operation is therefore central to our ability to infer the values of $\overline{L}_{A/B} = S_{A/B} Q_{A/B}$. We explain our protocol for doing this in section \ref{split}.

\begin{figure*}
\includegraphics[width=.99\textwidth]{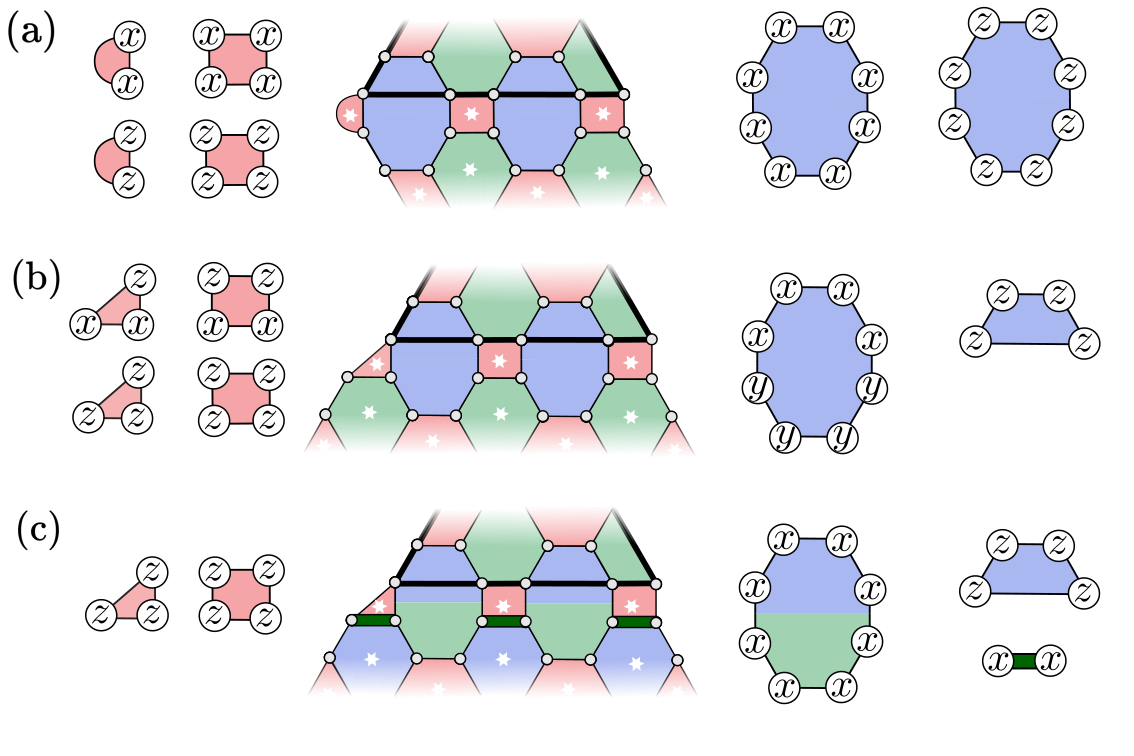}
\caption{Stabilizers connecting the ancilla system to data qubits. Different choices of stabilizer correspond to different logical measurements. Examples (a), (b) and~(c) correspond to case 1,2 and 3 in the main text. We have $\overline{P}^A_j = \overline{X}_j$ and $\overline{P}^B_j = \overline{Z}_j$ in~(a), $\overline{P}^A_j = \overline{P}^B_j = \overline{Z}_j$ in (b), and $\overline{P}^A_j = \openone$ and $\overline{P}^B_j = \overline{Z}_j$ in~(c). We can similarly measure $\overline{P}^A_j = \overline{X}_j$ and $\overline{P}^B_j = \openone$ by exchanging the Pauli-$X$ terms and Pauli-$Z$ terms in stabilizers in (c). Note that in general we will also encounter variations of the stabilizers shown which have the colors green and blue exchanged on the ancilla system. }
 \label{Fig:DomainWalls}
\end{figure*}

Next, we discuss the details of the local generators in $\mathcal{S}_M$ that let us perform these logical measurements fault-tolerantly.

Qubits in the bulk of the ancilla system and at the boundaries that are not merged with a data qubit are measured with standard color code stabilizers. 

The stabilizers in $\mathcal{S}_M$ connecting the ancilla system to each data qubit are determined by which of its Pauli operators are involved in the two logical measurements $\overline{L}_{A/B}$. Using the complete set of Clifford transversal gates on the data qubits, we can perform arbitrary logical measurements with just three different types of connections between the ancilla system and individual data qubits. 

Stabilizers that correspond to cases 1, 2 and 3, as defined previously, are shown in Fig.~\ref{Fig:DomainWalls} (a), (b) and (c) respectively. We can locally change the measurement basis of the examples shown in the figure using the transversal Clifford gate set of the triangular color code to obtain an arbitrary pair of Pauli measurements.  

As an aside, we remark that the different choices of stabilizer operators that connect data qubits and the ancilla can also be understood in terms of domain walls~\cite{FUCHS2002353, Bombin10, Kitaev12, 2017benpoking,2018litinskitwist, Barkeshli19, Bridgeman_2020, Bridgeman2020computingdatalevin} of the color code~\cite{Bombin_2011, Teo14, Yoshida15, Bridgeman17, Lavasani18, Kesselring_2018, gowda2021quantum} where the color code is viewed as a topologically ordered phase. We describe the novel types of domain walls that appear in this scheme in Appendix~\ref{App:SemiTransparent}. These domain walls are discussed in more detail in Ref.~\cite{kesselring2022anyon}.

\subsection{Splitting operation} \label{split}

During the splitting operation, we read out the ancilla system by measuring the local generators of $\mathcal{S}_{\textrm{anc}}$, projecting the ancilla system back onto a product state of Bell pairs between pairs of qubits that share a red edge. This disentangles the ancilla system and recovers the codespace of the data qubits. Since $\mathcal{S}_{\textrm{anc}}$ commutes with $Q_{A/B}$ and contains the red string operators $S_A$ and $S_B$, this ensures we can infer the values of $\overline{L}_{A/B} = \overline{Q}_{A/B} S_{A/B}$.

It is important to confirm that $\mathcal{S}_M$ does not contain any operators $\overline{L} U \in \mathcal{S}_M $ where $\overline{L} \not= \overline{L}_A,\, \overline{L}_B $ is a non-trivial logical operator and $U \in \mathcal{S}_{\textrm{anc}}$, such that we obtain the value of $\overline{L}$ during the splitting operation. Fig.~\ref{Fig:Architecture} gives an example based on the operator $\overline{Z}_2$ illustrating that this is the case. In general, we find that $U$ must consist of either blue or green string operators, or a mixture of both types, such that $U$ must anti commute with some red edge terms. It follows that $U \not\in \mathcal{S}$ and we therefore cannot infer the value of any $\overline{L} \not= \overline{L}_A,\, \overline{L}_B$.

\subsection{Comprehensive example}\label{example}

To summarise our scheme we provide an example in Fig.~\ref{Fig:Architecture}. This example illustrates each step of our protocol and shows a suitable ancilla system a for a given pair of measurements $\overline{L}_{A/B}$. We remark that the choice of $\overline{L}_{A/B}$ shown here contains examples of each of the three possible cases discussed previously and is thus comprehensive.

\subsection{Fault-tolerant implementation} \label{fault-tolerance}

At the level of implementing this logical operation with a realistic system we must be prepared to deal with the occurrence of measurement errors that occur when we transform it onto the merged system. A measurement error will mean a stabilizer reading will give an incorrect measurement outcome. Nevertheless, we will rely on random stabilizer readings to infer the value of the $Q_{A/B}$ operators. We must therefore repeat stabilizer readings to identify measurement errors. By repeating the stabilizer measurements $d$ times, we require that $\sim d / 2$ measurement errors must occur in order to incorrectly infer the value of $Q_{A/B}$. This is discussed in detail in Ref.~\cite{kesselring2022anyon}, where it is described how a syndrome is produced in $2+1$-dimensional spacetime. In such a setting we can use conventional decoding algorithms to identify Pauli errors and measurement errors alike. The use of $d$ repetitions of stabilizer generator readout at the merging step sets a timescale for the duration of a single merge operation. Let us caveat that here we are not considering circuit-level noise, and are thus not concerned with “hook” errors that might occur in syndrome extraction circuits and reduce the error correcting performance of the code. We address this in Sec.~\ref{Sec:circlevel}.

\begin{figure*}
\includegraphics[width=\textwidth]{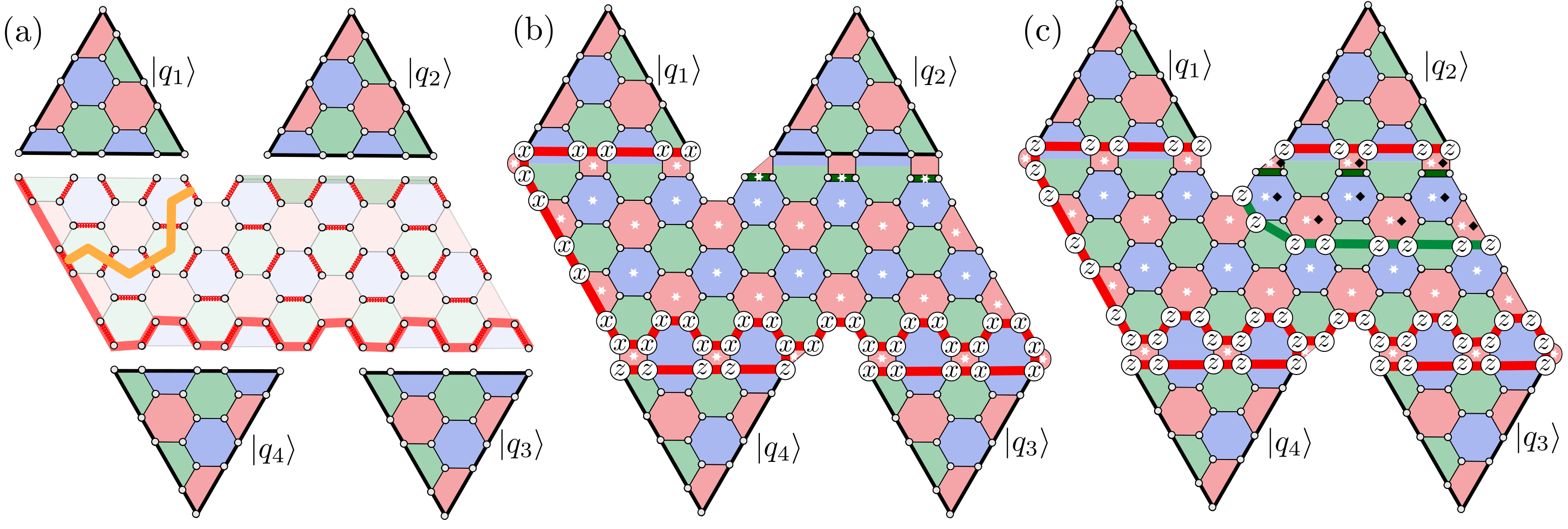}
\caption{Comprehensive example showing the measurement of two commuting logical operators $\overline{L}_{A/B}$ in our fault-tolerant color-code architecture. Here we have $N=4$ data qubits, labelled by state vectors $|q_j\rangle$ with $j=1,\ldots,4$ on which we measure the operators $ \overline{L}_A = \overline{X}_1 \overline{X}_3 \overline{Z}_4$ and $ \overline{L}_B = \overline{Z}_1 \overline{Z}_2 \overline{Z}_3  \overline{Z}_4$. \textbf{(a)} An ancilla system with a suitable configuration of boundaries, as described in the main text, is prepared in $\mathcal{S}_{\textrm{anc}}$, which is a product state of Bell pairs between pairs of qubits that share a red edge. $\mathcal{S}_{\textrm{anc}}$ contains the red string operators $S_A$ and $S_B$, defined in the main text, which are the product of red $X$ and $Z$ edge terms, respectively, a representative of which is shown in red. \textbf{(b)-(c)} During the merging operation we measure the new stabilizer group $\mathcal{S}_M$; see the main text for details. $\mathcal{S}_M$ anti-commutes with $S_A$ and $S_B$ but contains the operators $Q_A = \overline{L}_A S_A$ and $Q_B = \overline{L}_B S_B$, which are shown in red in (b) and (c) respectively and are the product of the stabilizer operators on faces marked by a white star. Repeating these stabilizer measurements $d$ times ensures fault-tolerance. During the splitting operation we then read out the ancilla system in $\mathcal{S}_{\textrm{anc}}$ and obtain the values of $S_A$ and $S_B$. One can check that they commute with the $Q$ operators and thus let us infer the values of $\overline{L}_A$ and $\overline{L}_B$. A string of errors acting on Bell pairs during the splitting operation can lead to a logical error. For instance an error on each of the Bell measurements that is crossed by the yellow line in (a) will lead to an incorrect reading of $S_A$ and/or $S_B$. This string has to be at least of length $d$ and so this part of the protocol is fault-tolerant and can be performed in constant time. To show that no unwanted logical information is measured during the splitting operation, we consider the example of $\overline{Z}_2$. The product of stabilizers of $\mathcal{S}_M$ marked by black diamonds in (c) gives operator $\overline{Z}_2 U$, with $U$ shown in green. It is easily checked that we do not infer the value of $\overline{Z}_2 $ as $U \not\in \mathcal{S}_{\textrm{anc}}$, since it anti-commutes with the red edge terms.}
\label{Fig:Architecture}
\end{figure*}

For practical application we must also consider the fault-tolerance of the readout of the red edge operators that we use to infer $S_{A/B}$. During the splitting operation, individual edge measurements give random outcomes, constrained such that the product of the red edges about the boundary of a blue or green face are equal to the value of the corresponding stabilizer at that face. Similarly, for stabilizers supported on both the data qubits and ancilla system, edges may be constrained to give a weight-eight stabilizer when multiplied together with a weight-four boundary stabilizer of $\mathcal{S}_\textrm{data}$. An error experienced during a Bell measurement from $\mathcal{S}_\textrm{anc}$ will therefore violate its two adjacent stabilizers. Errors on the Bell pairs then appear as strings that run dual to the red edges where violated stabilizers can be measured at their end points. We can correct these errors using a matching decoder to pair violated stabilizers. A logical error will occur if a string runs between two distinct boundaries of the ancilla system such that we obtain the incorrect value of $S_{A/B}$; an example of such an error is shown in Fig.~\ref{Fig:Architecture}. We can incorporate this error correction procedure into the practical situation where measurements are unreliable by comparing the inferred values of stabilizer operators to earlier readings of the stabilizer measurements to identify stabilizer violations~\cite{Dennis02}.

\section{Space-time overhead estimates}\label{Sec:spacetime}

We now estimate the overhead needed by our scheme to achieve a given code distance $d$. The code distance gives a leading order approximation of the logical failure rate, $\sim p^{d/2}$, where error events occur with low probability $p$ assuming a suitable stabilizer readout circuit and a decoder that can correct errors consisting of $\lesssim d /2 $ error events.

To estimate the overhead, we consider running algorithms where magic states are injected using logical Pauli measurements, grouping measurements into mutually commuting layers; see~\cite{Litinski_2019}. We compare our results to a space-time efficient surface-code architecture, the \emph{fast data block scheme} of Ref.~\cite{Litinski_2019}.  This scheme, like the color-code scheme we propose, enjoys a low space-time resource cost, largely because the time spent performing single-qubit Clifford operations on data qubits is negligible. Fundamentally, this saving is due to the fact that in both schemes, the data qubits have all of their logical operators supported on qubits that are adjacent to the ancilla system.

Our results are shown in Table \ref{tab:my_label}.  For an explicit example with $N=100$ logical data qubits, our color-code scheme offers a $ 3.1\times$ reduction in space-time overhead compared to the fast data block scheme.

\begin{table}
    \centering
\begin{tabular}{|c|c|c|c|}
\hline
scheme & space ($d^2$) & time ($d$) & spacetime ($d^3$) \\
 \hline
 color code & $1.5N$ & $0.5T$ & $0.75NT$ \\
 surface code  & $ 2N+\sqrt{8N}+1$ & $T$ & $(2N+\sqrt{8N}+1)T$ \\
 \hline
\end{tabular}
    \caption{Overhead of our scheme compared with the surface-code-based fast data block~\cite{Litinski_2019} for $N$ logical data qubits where the depth-$T$ circuit is composed of long sequences of commuting measurements.
    The space cost is $\frac{3}{4} d^2$ physical qubits per logical qubit for the color code on the hexagonal lattice~\cite{landahl2011faulttolerant}. The ancilla system also requires $\frac{3}{4}d^2$ physical qubits per logical data qubit giving the total space cost of the color-code scheme to leading order.  Time cost improvements originate from measuring pairs of commuting logical Pauli operators in parallel.
    }
    \label{tab:my_label}
\end{table}

The factor-of-two reduction in temporal overhead with our scheme arises from the parallelism of performing two arbitrary commuting logical Pauli measurements simultaneously.  This assumes that we can measure a complete generating set of stabilizers per unit time, independent of the depth of the syndrome readout circuit.  While color code extraction circuits are more complex than those of the surface code, this assumption is nonetheless well motivated in experimental platforms where the syndrome readout circuit time is dominated by qubit measurement and reset rather than unitary gates. We assume that a distance $d$ against measurement errors can be achieved by repeating the syndrome readout circuit $d$ times.

The origin of the qubit count (space) overhead savings is twofold. Firstly, physical qubits are spared because the color code has a lower qubit overhead compared to a surface code with an equivalent code distance~\cite{2007bombincomparative}. Secondly, the width of the boundary shared between the ancilla and a data qubit is twice as large for the fast data block scheme with the surface code as compared with our color-code scheme.

The resource cost of a basic implementation of magic state distillation using our color-code scheme is also competitive with optimised surface-code schemes.  For the surface code, a 15-to-1 magic state distillation protocol can exploit the fact that the logical measurements are made in the same basis to achieve a very low spatial overhead~\cite{Litinski_2019,2019litinskinotascostly}. As such, our color-code scheme would offer only a modest improvement in terms of spatial resources ($10.5d^2$ compared with $11d^2$). It would nonetheless offer a significant $1.8\times$ reduction in the time cost of the required sequence of 11 commuting measurements, leading to an overall $1.9\times$ improvement in space-time overhead for magic state distillation.  We note that magic state distillation also requires the preparation of high-fidelity magic states; methods for the efficient preparation of high-fidelity magic states using color codes have recently been shown in Ref.~\cite{chamb2020} and will complement our scheme.

\begin{figure}
\centering
\includegraphics[width=1.0\linewidth]{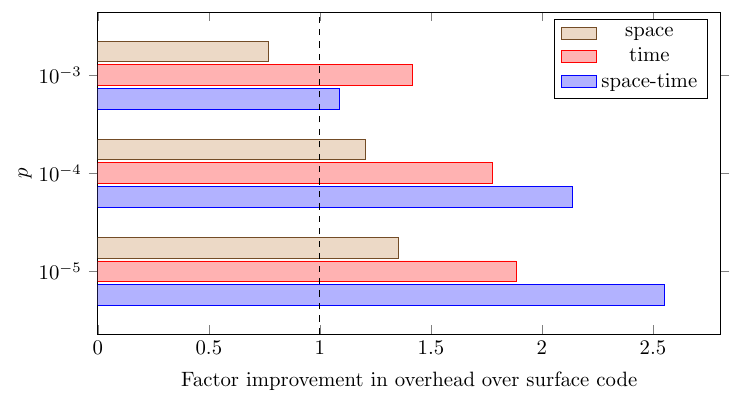}
\caption{Improvement in space-time overhead of our color-code architecture over the surface code for a $N=100$ qubit, $10^8$ T-gate computation under circuit level noise for different physical error rates. The color-code architecture is faster and more space-time efficient at any error rate and begins to use fewer physical qubits at a physical error rate of $\sim4\times10^{-4}$.}
\label{plots}
\end{figure}

\section{Performance under circuit-level noise}\label{Sec:circlevel}

While the code distance can serve as a good proxy for the error correcting properties of a code, in practice we are interested in a target logical failure rate. The logical failure rate will be impacted by the performance of the decoder as well as properties of the code. In general, the additional complexity of syndrome extraction with the color code means that it tends to perform worse than the surface code in terms of error-correction threshold under circuit level noise~\cite{2012surface}. To account for this, we follow Ref.~\cite{Litinski_2019} and compare the overheads of our scheme and the fast data block scheme under circuit-level noise for a typical algorithm~\cite{Babbush_2018} with 100 qubits and a T-count of $10^8$. This equates to $10^8$ logical Pauli measurements assuming encoded magic states at the required logical error rate are available. Our results are shown in Fig.~\ref{plots}. We find that the color-code architecture is the more space-time efficient choice, achieving a $10\%$ lower space-time overhead for a physical error rate of $10^{-3}$. While at this error rate the color code uses $30\%$ more qubits, remarkably it starts outperforming the surface code on this metric at error rates around $4\times 10^{-4}$ or lower. At physical error rates lower than $10^{-4}$, the space-time overhead of the color-code architecture is more than two times lower than that of the surface code.

We fix the total chance of failure of the algorithm due to a logical error to be $1\%$ and estimate $p_L$, the logical error rate per code cycle for a logical surface code qubit with distance $d_s$ under circuit level noise parameterized by $p$, as $p_{L}(p, d_s)=0.1(100 p)^{(d_s+1) / 2}$~\cite{fowler2019low}. To compute the color code distance $d_c$ required to achieve the same logical error rate as the surface code with distance $d_s$ under circuit level noise with rate $p$, we use the approximation $d_{c}=d_{s}\bigl(\frac{\log p / p_{\mathrm{th}}^{(s)}}{\log p / p_{\mathrm{th}}^{(c)}}\bigr)$~\cite{landahl2014quantum} with circuit noise threshold values $p_{\mathrm{th}}^{(c)} = 0.37\%$~\cite{beverland2021cost} and $p_{\mathrm{th}}^{(s)} = 0.67\%$ ~\cite{Stephens_2014}. We remark that these assumptions should be considered to be conservative given recent results on improved syndrome extraction circuits and decoders \cite{gidney2023new,zhang2023facilitating} for the color code and that in the near future the gap between color code and surface code performance under circuit-level noise might be closed entirely, such that the full extent of the resource savings discussed in Sec.~\ref{Sec:spacetime} may be realized in practice.

\begin{figure}
\centering
\includegraphics[width=0.8\linewidth]{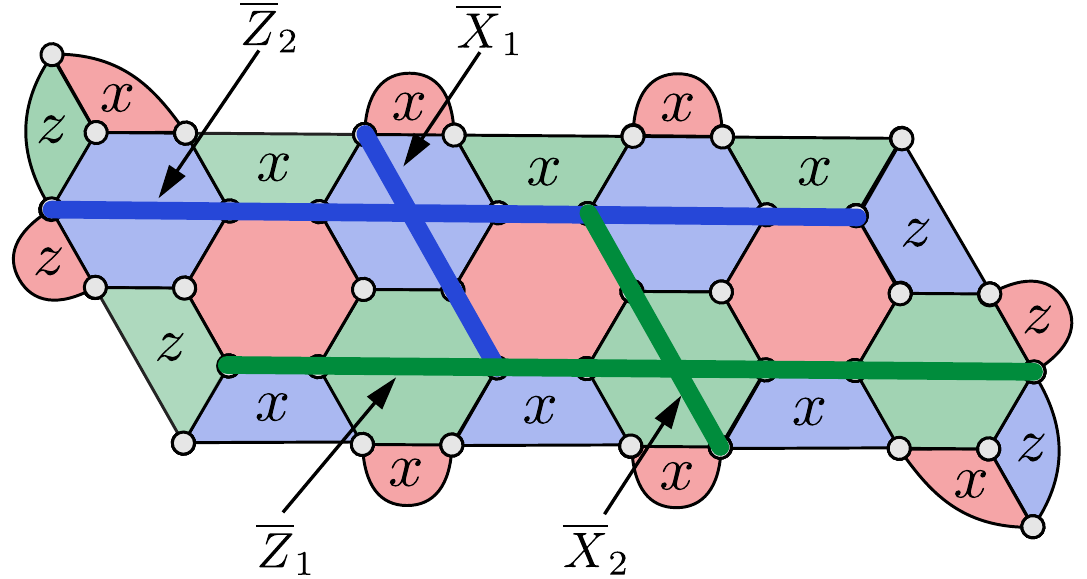}
\caption{The thin color code encodes two logical qubits. Faces marked with an $x$ ($z$) only support a single Pauli-$X$ (Pauli-$Z$) type stabilizer. We replace its colored boundaries with Pauli boundaries~\cite{Kesselring_2018} such that its least-weight logical operator of physical Pauli-$X$ (Pauli-$Z$) terms has weight $d_x=3$ ($d_z=7$). The ratio $d_z /d_x$ is chosen to minimize the logical error rate.}
\label{thin}
\end{figure}

\section{Discussion}\label{Sec:Discussion}

The color code has the potential to offer significant reductions in the resource cost of fault-tolerant quantum computing in two dimensions.  Other recent developments potentially offer additional resource savings, such as reducing time overhead using Pauli-Y stabilizer measurements~\cite{sahay2021decoder} (see also Ref.~\cite{delfosse2020short}), and by tailoring architectures to common types of structured noise~\cite{Aliferis08, Li19, chamberland2020building, pablo2021, xzzx2021}, for example by using a thin color code, shown in Fig.~\ref{thin}. Importantly, using Pauli-type boundaries~\cite{Kesselring_2018}, we can vary the distance of the thin color code against bit- and phase-flip errors independently to optimise the logical failure rate. Our work therefore motivates the development of better decoders and syndrome extraction circuits for the color code, where there is clear potential for further gains in both performance and efficiency. We may also find that we can improve resource scaling using other color code lattice geometries~\cite{landahl2011faulttolerant, kubica2019efficient}.

Our results highlight the role of domain walls in allowing parallel measurement of an arbitrary pair of commuting Pauli operators. A general theory of domain walls may reveal new objects, allowing for further reductions in resource cost. For instance, while we have focused on measuring two arbitrary commuting Pauli operators, in general we can measure a larger number of commuting measurements that respect certain physical constraints by subdividing the ancilla system using additional domain walls. We will also benefit from more general methods of compiling topological circuits~\cite{Litinski_2019, 2019gidneymagic, deBeaudrap2020zxcalculusis} using the extensive toolbox of topological defects available to the color code. We leave the development of compilation methods for color code quantum computing as well as a general theory of domain walls to future work.

\begin{acknowledgements}
We are grateful for comments and conversations with D. Barter, J. Bridgeman, J. Eisert, C. Gidney, A. Landahl, D. Litinski, J. Magdalena, M. Newman and M. Vasmer. This work is supported by the Australian Research Council via the Centre of Excellence in Engineered Quantum Systems (EQUS) project number CE170100009, and by the ARO under Grant Number: W911NF-21-1-0007. FT is supported by the Sydney Quantum Academy. MSK is supported by the FQXI, the BMBF (RealistiQ), the Munich Quantum Valley (K8), the Einstein Research Unit, and the DFG (CRC183-B01).
\end{acknowledgements}

\appendix

\section{Semi-transparent domain walls in the color code}

\label{App:SemiTransparent}

In the main text, the color code was used as a quantum error-correcting code.  To perform fault-tolerant logical measurements, lattice surgery was implemented by measuring a new stabilizer group of a merged code. Different choices of these stabilizers connecting the ancilla system to the data qubits give rise to different logical measurements. In order to capture the physics of the different stabilizers that make this connection during lattice surgery protocols, we turn to the theory of domain walls. More generally, we use {\it semi-transparent} domain walls to perform an arbitrary pair of commuting measurements. This appendix summarises some key details of semi-transparent domain walls that we use in this work. A comprehensive discussion on semi-transparent domain walls in the color code is given in~\cite{kesselring2022anyon}.

Here, after briefly reviewing the anyons, boundaries and transparent domain walls of the color code, we focus on a specific semi-transparent domain wall already encountered in the main text. This example serves to illustrate the essential physics before we generalise to describe all semi-transparent domain walls of the color code phase. Finally, we study the semi-transparent domain walls in the unfolded picture.

We view the color code as a topological lattice model by defining a Hamiltonian $H = - \sum_j S_j$, where $\{S_j\}$ are a set of local stabilizer generators of the color code.  The code space then gives the ground state of this Hamiltonian, and we can consider errors as operators that create excitations.
We follow the notation of Ref.~\cite{Kesselring_2018} to describe the anyonic excitations (or simply anyons) of the color code, briefly summarised here. Faces that support violated stabilizer operators are occupied by anyons. There are different species of these quasiparticle excitations, each with distinct properties. We concentrate on the bosonic anyons whose species have two labels; a color label and a Pauli label. The color of an anyon is determined by the face operator on which it is supported; red, green or blue, labelled $r$, $g$ or $b$.
Each Pauli operator creates a different type of anyon that we distinguish using a label $x$, $y$ or $z$. Bosonic anyons take one color label and one Pauli label to obtain nine bosonic anyons. The complete anyon model of the color code also includes six fermionic anyons; we do not consider them explicitly here, but note that they are obtained by taking composites of the bosonic excitations and as such, we implicitly deal with fermionic excitations as well. See~\cite{Kesselring_2018} for more details.

The bosonic anyons of the color code can be arranged in a $3\times3$-grid as shown in Table~\ref{tab:CC3x3anyons}. The grid is arranged to capture the data of the anyon model. This contains self-exchange and braid statistics between different anyons as well as their fusion rules. All shown anyons are bosons, i.e. have trivial self-exchange statistics. Two anyons in the same column, or in the same row, braid trivially with one another. A braid between any other pair of anyons in the grid gives rise to a non-trivial global phase. Two bosons in the same row (column) fuse to the third boson in said row (column), fusion between bosons of different rows and columns creates one of six fermions.

\begin{table}[h]
	\begin{tabular}{ c | c | c }
         $rx$ & $gx$ & $bx$ \\ \hline    
         $ry$ & $gy$ & $by$ \\ \hline    
         $rz$ & $gz$ & $bz$ 
    \end{tabular}
	\caption{
		The nine bosonic anyons of the color code can be labeled using a color label and a Pauli label.
		The three colors, $r$, $g$, and $b$ are give in the columns, the Pauli labels $x$, $y$, and $z$ are associated with rows.
		This arrangement encodes the self-exchange and braid statistics as well as the fusion rules of the color code anyons completely.
	}
	\label{tab:CC3x3anyons}
\end{table}

Ref.~\cite{Kesselring_2018} concentrates on two types of objects for the color code phase, {\it gapped boundaries}~\cite{Levin13,Barkeshli13} and {\it transparent domain walls}~\cite{Kitaev12}. The theory of {\it semi-transparent domain walls} generalises the aforementioned objects. Let us briefly review gapped boundaries and transparent domain walls before describing the more general case.

If an anyon is brought close to a boundary, there are two possible mechanisms we can observe: condensation or confinement. We can regard a boundary as a mode that can absorb certain types of anyon, where the state of this mode is determined by the anyons that occupy the boundary.  Condensation means that an anyon is absorbed by the boundary, thereby changing the state of the boundary. The mass of the quasiparticle is in turn dissipated. Similarly, individual anyons can be created locally out of a boundary from which it can be condensed. In contrast, a boundary might confine certain types of anyon, meaning that the anyon cannot be absorbed by the boundary. It therefore remains localised close to the boundary.

There are certain rules a boundary must respect in terms of the anyons it can condense or confine in order for it to maintain the energy gap of the Hamiltonian under local perturbations~\cite{Levin13}. It can be shown that the set of all of the anyons that can be condensed by a gapped boundary must a) only contain bosons, b) be closed under fusion and that c) any pair of condensable anyons must braid trivially with one another. In Ref.~\cite{Kesselring_2018} it was shown that a gapped boundary of the color code can condense three different types of bosonic anyon, all sharing either a row or a column of Table~\ref{tab:CC3x3anyons}. The remaining anyons are necessarily confined at the boundary. Given that the table has three rows and three columns, there exist six different types of gapped boundary for the color code.

Unlike boundaries, transparent domain walls are found in the bulk of the lattice. Anyons that traverse a domain wall might get transformed to an anyon of another type. For the case of transparent domain walls, the mapping describing the transformation of all anyons across the domain wall is an automorphism, in the sense that the properties of the anyons (like fusion, self-exchange and braid statistics) are unchanged upon undergoing the mapping.  Using the structure of Table~\ref{tab:CC3x3anyons}, Ref.~\cite{Kesselring_2018} argued that the automorphisms that are permitted by a transparent domain wall should preserve the rows and columns of the grid.
The allowed automorphisms therefore correspond to permutations of the rows and the columns of Table~\ref{tab:CC3x3anyons}. The mapping may also transpose the table about its leading diagonal. All together, we discover $ 72$ different transparent domain walls~\cite{Yoshida15,Kesselring_2018}.

\begin{figure}
\includegraphics[width=1.00\linewidth]{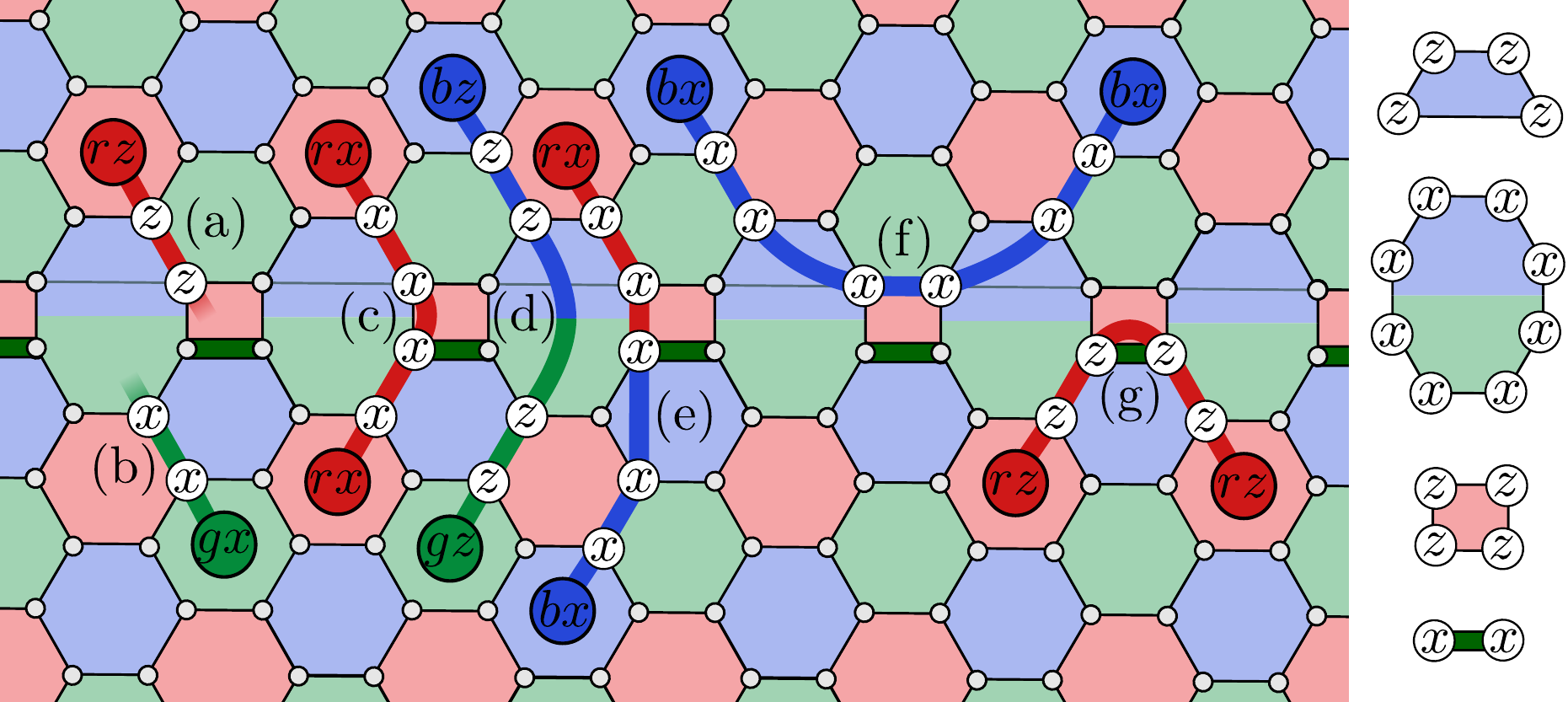}
\caption{
The action of the semi-transparent domain wall presented in Fig.~\ref{Fig:DomainWalls}(c) on anyonic excitations of the color code.
(a)~and (b)~show a single excitation that is condensed at the semi-transparent domain wall.
Note, the condensing anyons might be different depending on the direction from which they approach the domain wall.
(c),~(d)~and (e)~show anyons that are transmitted over the domain wall. Note how processes (c)~and (e)~are related by multiplication with process (b).
(f)~and (g)~show anyons that are confined on each side of the domain wall.
}
\label{Fig:AnyonsSTDW}
\end{figure}

A semi-transparent domain wall demonstrates the behaviour of both boundaries and domain walls. A semi-transparent domain wall is supported on a line of the lattice. Anyons that approach the domain wall may condense, they may confine, or they may be transmitted, while also undergoing some mapping that respects their mutual braid statistics. We show an example of one such semi-transparent domain wall in Fig.~\ref{Fig:AnyonsSTDW} together with all of the distinct mechanisms that anyonic excitations can undergo at the domain wall. The example is taken from Fig.~\ref{Fig:DomainWalls}(c) in the main text.

\begin{table}[h]
	\begin{tabular}{ c | c | c }
         $A$ & $D$ & $D$ \\ \hline    
         $A$ & $D$ & $D$ \\ \hline    
         $C$ & $B$ & $B$ 
    \end{tabular}
	\begin{tabular}{ c }
         $|$ \\    
         $|$ \\    
         $|$ \\     
         $|$ 
    \end{tabular}
	\begin{tabular}{ c | c | c }
         $A$ & $C$ & $A$ \\ \hline    
         $D$ & $B$ & $D$ \\ \hline    
         $D$ & $B$ & $D$ 
    \end{tabular}
	\caption{
		Different semi-transparent domain walls can be expressed using two grids of bosonic anyons.
		Anyons with the labels $A$ and $B$ can pass through the domain wall, turning into one of the anyons with the same label ($A$ or $B$) on the other side.
		Anyons labelled $C$ condense when approaching the domain wall.
		The other four anyons, labelled $D$, get confined at the domain wall.
	}
	\label{tab:STDWanyons}
\end{table}

The full classification of semi-transparent domain walls in the color code phase can again be understood using the grid. Specifically, we take two copies of the grid, one corresponding to the `left-hand side' of the domain wall and the other corresponding to the `right-hand side', as shown in Table~\ref{tab:STDWanyons}. 

On, each side of the domain wall a single arbitrary choice of boson can be condensed. We mark this with a $C$ label on the left- and right-hand side in Table~\ref{tab:STDWanyons}.

Given an anyon that can condense at the domain wall, it follows that all bosonic anyons that braid non-trivially with the condensed anyon must be confined at the domain wall using generalisations of arguments that are presented in~\cite{Kesselring_2018}. These are the four bosonic anyons that share neither a row nor a column with the anyon marked $C$. We mark these with the letter $D$ on both the left- and right-hand grid. 

The remaining anyons that braid trivially with anyon $C$ can be transmitted across the semi-transparent domain wall. We label these remaining anyons $A$ and $B$. Upon transmission, the $A$($B$) labels of the left-hand grid are mapped onto the $A$($B$) labels of the right-hand grid. 

Note that there are two anyons labeled $A$ and two labeled $B$. We do not distinguish between these charges because at the semi transparent domain wall they are equivalent up to fusion with the charge that is condensed. Given that we can create the condensed charges locally at the domain wall, and that the fusion between two anyons in the same row or column give the third anyon of the respective row or column, the two charges labeled $A$ and the two charges labeled $B$ are in effect equivalent near to the domain wall.

Following these rules we find the total number of semi-transparent domain walls. First of all, we can choose the condensing anyon, $C$, on both the left-hand side and right-hand side arbitrarily. This fixes the confined charges on each grid.  Without loss of generality, let us also fix the $A$ and $B$ labels on the left-hand grid. We have one final degree of freedom, namely, which charges the $A$ labels map to on the right-hand grid and, likewise, which charges the $B$ labels map to on the right-hand grid. There are two possible choices to configure the $A$ and $B$ labels on right-hand grid.

Given the nine choices of condensing anyons on the left- and right-hand side grid, we obtain $81$ semi-transparent domain walls. Then, together with the binary choice for how to configure the $A$ and $B$ labels on the right-hand grid we arrive at $81 \times 2 = 162$ semi-transparent domain walls.

For completeness, we can also obtain 36 opaque domain walls by choosing any configuration of two of the six gapped boundaries at either side of a vacuum phase, i.e., the gap between two boundaries where only the trivial vacuum anyon can exist.
These opaque domain walls are used in the lattice surgery protocol presented in the main text where data qubits do not support either of the two logical measurements on a given step of the algorithm.

\begin{figure}[tb]
\centering
\includegraphics[width=1.00\linewidth]{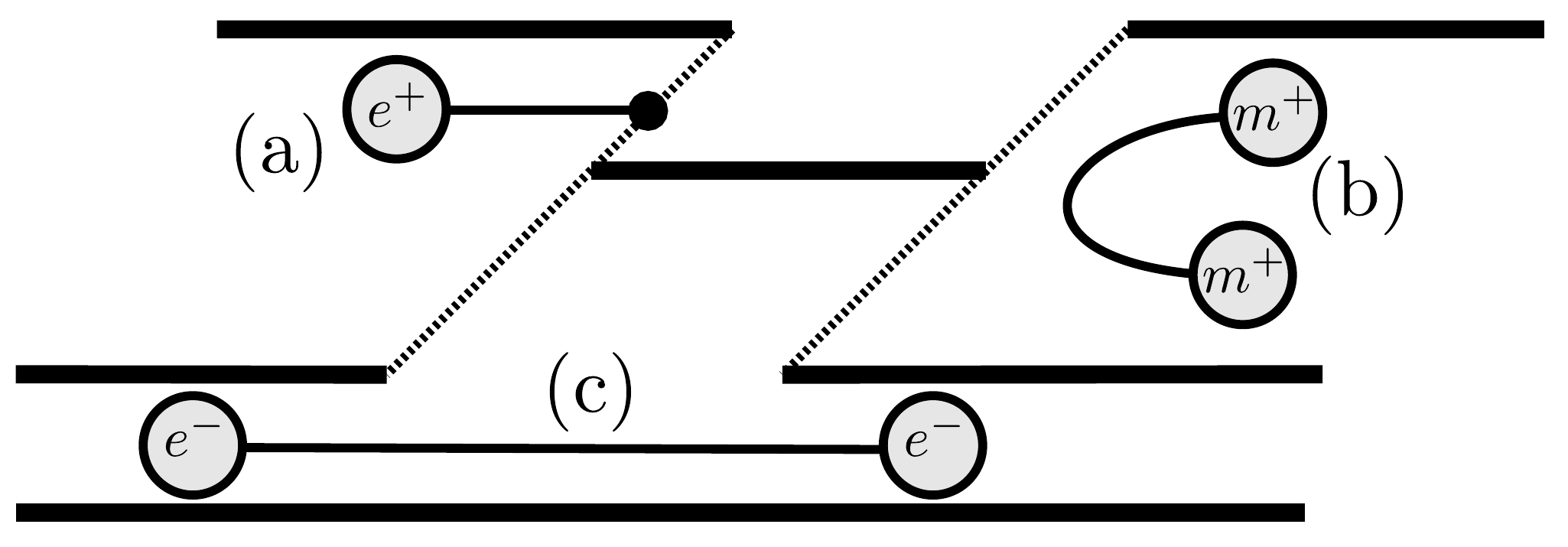}
\caption{
The color code corresponds to two stacked copies of the toric code.
Its semi transparent domain walls can be understood in this picture.
While one layer is interrupted by boundaries, the other layer connects the two sides.
(a) shows an anyon condensing when approaching the domain wall from the left.
(b) shows an anyon confined on the right of the domain wall.
(c) shows an anyon passing through the domain wall on the bottom layer.
}
\label{Fig:DWunfolding}
\end{figure}

We can understand the physics of the semi-transparent domain walls of the color code by unfolding~\cite{Bombin_2012,Kubica_2015, Kesselring_2018} it into two copies of the toric code.
As shown in Fig.~\ref{Fig:DWunfolding}, we terminate the top layer on both sides with a rough boundary while the bottom layer connects through the domain wall.
Anyons on the bottom layer can pass through the domain wall, while anyons on the top layer will either condense or confine depending on their species.
We redundantly obtain all possible semi-transparent domain walls from this simple example by adding transparent domain walls to either side of the displayed semi-transparent domain wall.
The unfolded color code domain walls are described in Ref.~\cite{Kesselring_2018}.

\providecommand{\noopsort}[1]{}\providecommand{\singleletter}[1]{#1}%

\end{document}